\documentclass[showpacs,pra,twocolumn]{revtex4-1}
\usepackage{epsfig,amsmath}
\usepackage{subfigure}
\usepackage{graphicx}
\usepackage{dcolumn}
\usepackage{stmaryrd}
\usepackage{mathrsfs}
\usepackage{pifont}
\usepackage{amsthm}
\usepackage{amssymb}
\usepackage{bm}
\usepackage{latexsym}
\usepackage{hyperref}
\usepackage{color}
\usepackage{appendix}

\newcommand{\beq}{\begin{equation}}
	\newcommand{\eeq}{\end{equation}}
\newcommand{\beqa}{\begin{eqnarray}}
	\newcommand{\eeqa}{\end{eqnarray}}

\newcommand{\lam}{\lambda}

\begin{document}
\author{Zhao-Ming Wang$^{1,3}$}
\altaffiliation{ These authors contributed to the work equally and
should be regarded as co-first authors}
\author{S. L. Wu$^{2,3}$}
\altaffiliation{ These authors contributed to the work equally and
should be regarded as co-first authors.}
\author{Mark S. Byrd$^{4}$}
\email{mbyrd@siu.edu }
\author{Lian-Ao Wu$^{3,5,6}$}
\email{lianao.wu@ehu.es}
\affiliation{$^{1}$ College of Physics and Optoelectronic Engineering, Ocean University of China, Qingdao 266100, China \\
$^{2}$ School of Physics and Materials Engineering, Dalian Nationalities University, Dalian 116600 China\\
$^{3}$ Department of Physics, University of the Basque Country UPV/EHU, 48080 Bilbao, Spain\\
$^{4}$ Department of Physics, Southern Illinois University, Carbondale, Illinois 62901-4401, USA \\
$^{5}$ IKERBASQUE, Basque Foundation for Science, Bilbao, 48011, Spain\\
$^{6}$ EHU Quantum Center, University of the Basque Country UPV/EHU, Leioa, Biscay 48940, Spain }
\title{Going beyond quantum Markovianity and back to reality: An exact master equation study}
\date{\today}

\begin{abstract}
The precise characterization of dynamics in open quantum systems often presents significant challenges, leading to the introduction of various approximations to simplify a model. One commonly used strategy involves Markovian approximations, assuming a memoryless environment. In this study, such approximations are not used and an analytical dynamical depiction of an open quantum system is provided.  The system under consideration is an oscillator that is surrounded by a bath of oscillators. The resulting dynamics are characterized by a second-order complex coefficient linear differential equation, which may be either homogeneous or inhomogeneous. Moreover, distinct dynamical regions emerge, depending on certain parameter values. Notably, the steady-state average excitation number (AEN) of the system shows rapid escalation with increasing non-Markovianity, reflecting the intricacies of real-world dynamics. In cases where there is detuning between the system frequency and the environmental central frequency within a non-Markovian regime, the AEN maintains its initial value for an extended period. Furthermore, the application of pulse control can effectively protect the quantum system from decoherence effects without using approximations.  The pulse control can not only prolong the relaxation time of the oscillator, but can also be used to speed up the relaxation process, depending on the specifications of the pulse.  By employing a kick pulse, the Mpemba effect can be observed in the non-Markovian regime in a surprisingly super-cooling-like effect.
\end{abstract}

\maketitle

\section{Introduction} 

An open quantum system is a system in contact with a surrounding reservoir (bath), where interactions and possible correlations between the system and reservoir persist throughout the evolution. Various method have been used to describe the intricate dynamics inherent in such systems \cite{Breuer}. Typically, the reservoir is modeled as with infinite degrees of freedom, and the focal point lies in the reduced density matrix of the system, acquired through the tracing out of the reservoir degrees of freedom. However, during this procedure, the information related to the correlations becomes obscured, rendering the derivation of a deterministic differential equation of motion challenging without resorting to approximations. Consequently, various approximations have been invoked to address these correlations and corresponding dynamics.

The so-called Lindblad-Markovian master equation stands out as a straightforward method for characterizing an open system, involving the assumption of Markovian approximation and the complete neglect of memory effects within the bath \cite{Kossakowski1976,Lindblad1976}.  The investigation pertains to the Born-Markov-approximated Redfield quantum master equation description for an open system composed of noninteracting particles weakly connected to multiple reservoirs characterized by different temperatures and chemical potentials \cite{Archak}. The derived Redfield equation undergoes simplification to yield the master equation through additional approximations. The applicability of the Markovian approximation is limited to short correlation times. Conversely, when considering scenarios involving memory effects, such as strong couplings or low-temperature environments, a non-Markovian description becomes imperative.  For example, the problem of a particle in contact with a reservoir of oscillators can give rise to a master equation for the motion of the particle when one traces over the reservoir degrees of freedom \cite{Hu}. A non-Markovian process can be incorporated into a Markovian framework through the introduction of a time-dependent Lindblad generator on an extended state space, facilitating the establishment of genuine quantum trajectories \cite{Breuer2004}. Various master equations, including the axiomatic Lindblad-Gorini-Kossakowski-Sudarshan formulation \cite{Kossakowski}, the phenomenological post-Markovian counterpart \cite{Shabani}, and the Stochastic Schr\"odinger equation \cite{Semina1,Semina2}, have undergone extensive investigation. These methodologies, have proven useful in quantum optics \cite{Scully}, quantum computation \cite{Verstraete,Kliesch}, condensed matter physics \cite{Morigi}, and the theory of decoherence \cite{Zurek}, and have broad applicability.

Langevin equations have proven effective in describing the interaction between classical systems and complex thermal reservoirs characterized by linear dissipation, employing simple terms such as stochastic forces and memory friction \cite{Benguria,Oliveira,Olavo}. The macroscopic depiction of a quantum particle experiencing dissipation within an arbitrary external potential can be described using the generalized Langevin equation \cite{Ford}. However, an exact equation of motion for the general case in the quantum realm remains unavailable.  While the Markovian approximation has established Langevin equations \cite{Hanggi}, this paper delves into the derivation and discussion of dynamics within open quantum systems, accommodating arbitrary strength and correlation time scales.  Without resorting to any approximations, the relaxation dynamics of a single-mode harmonic oscillator embedded in a reservoir of independent bosonic oscillators is solved. Notably, an analytical solution is obtained for the system dynamics, providing a profound insight into the microcosm of quantum phenomena. Further, it is shown that the average value of the system's AEN can be perfectly preserved when implementing successive external pulses on the target system. The influence of important pulse parameters on the effectiveness of pulse control are investigated.  It is shown that a non-Markovian Mpemba effect can be induced by a ``kick pulse". When the non-Markovianity is weak, the pulse on the initial state with low temperature will prolong the relaxation time. In the strong non-Markovian regime, the relaxation process is sped up by applying a selected kick pulse. This is the Mpemba effect mentioned earlier.

\section{Open system dynamics}

Consider a system featuring a single-mode harmonic oscillator with frequency $\omega_0$ coupling to a reservoir comprised of independent bosonic oscillators. For convenience, we consider a bilinear exactly-solvable model,
\begin{eqnarray}
H=\omega_0 a^\dagger a+\sum_{j}\omega_j b_j^\dagger b_j+\sum_j( g_ja^\dagger b_j+ g_j a b_j^\dagger),
\end{eqnarray}
where $g_j$ denotes the coupling strength between the oscillator and the $j$th mode of the reservoir. The operators $a_j$ and $a_j^\dagger$ represent the system creation and annihilation operators, respectively, with frequency $\omega_0$.  Similarly,
$b_j$ and $b_j^\dagger$ represent the reservoir creation and annihilation operators, respectively, of the reservoir modes with frequencies $\omega_j$.  In this paper, we consistently employ the convention $\hbar=1$. The Heisenberg equations of motion are given by
\begin{eqnarray}
\dot{a}=-i\omega_0 a-i\sum_{j}g_jb_j,
\label{eq2}\
\dot{b_j}=-i\omega_j b_j-ig_j^* a.
\label{eq3}
\end{eqnarray}
Upon formal integration of Eq.~(\ref{eq3}), the equation of motion for the reservoir operator is
\begin{equation}
b_j(t)=b_j(0)e^{-i\omega_j t}-ig_j^*\int_{0}^{t}dsa(s)e^{-i\omega_j(t-s)}.
\label{eq4}
\end{equation}
Inserting (\ref{eq4}) back into (\ref{eq2}), it yields
\begin{equation}
	\frac{ da(t)}{dt}=-i\omega_0 a(t)+F(t)-\int_{0}^{t}ds G(t-s) a(s),
	\label{eq6}
\end{equation}
with $F(t)=-i\sum_{j} g_j b_j(0)e^{-i\omega_j t}$, $G(t-s)=\int d\omega J(\omega)e^{-i\omega (t-s)}$. The integro-differential equation shows that the model Hamiltonian can be exactly solved numerically. The model can also be extended to include another kinds of bilinear terms like $ab_j$ and $ b_ib_j$ with a similar solution. Furthermore, it is notable that we can obtain a fully analytical solution if we opt for the Lorentzian spectral density, characterized by a positive peak frequency and width. The spectral density is defined as $$J(\omega)=\sum_{j}\lvert g_j \rvert^2 \delta(\omega-\omega_j)=\frac{\Gamma \gamma^2/2\pi}{(\omega-\Omega)^2+\gamma^2}.$$  Here, $\gamma^{-1}$ represents the memory time, $\Omega$ denotes the central frequency of the environment, and $\Gamma$ serves as the global dissipation rate. Analytical calculation yields the correlation function $G(t-s)=\frac{1}{2}\Gamma \gamma e^{-(\gamma+i \Omega) \lvert t-s \rvert}$.

We represent the annihilation operator of the harmonic oscillator as a linear combination of the initial annihilation operators for both the system and the reservoir, given by the equation
\begin{eqnarray}
a(t)=A(t)a(0)+\sum_{j} B_j(t) b_j(0).
\label{eq10}
\end{eqnarray}
The expansion coefficients $A(t)$ and $B_j(t)$ satisfy the initial conditions $A(0)=1$ and $B_j(0)=0$. Substituting Eq.~(\ref{eq10}) into Eq.~(\ref{eq6}), we derive the equations governing the expansion coefficients:
\begin{eqnarray}
&\dot{A}(t)=-i\omega_0 A(t)-\int_0^tds G(t-s)A(s),
\label{eq11} \\
&\dot{B_j}(t)=-i\omega_0 B_j(t)-\int_0^tds G(t-s)B_j(s)-ig_je^{-i\omega_j t}.
\label{eq12} \nonumber\\
\end{eqnarray}
These equations come with the initial conditions $\dot{A}(t=0)=-i\omega_0$ and $\dot{B_j}(t=0)=-ig_j$. For the correlation function $G(t-s)$, the derivative is expressed as $\dot{G}(t-s)=-(\gamma+i\Omega) G(t-s)$. Further differentiating Eq.~(\ref{eq11}) or Eq.~(\ref{eq12}) with respect to time, we obtain
\begin{eqnarray}
&&\ddot{A}(t)+(a+ib) \dot{A}(t)+(c+id) A(t)=0.
\label{eq141} \\
&&\ddot{B}_j(t)+(a+ib) \dot{B}_j(t)+(c+id) B_j(t)\label{eq232} \nonumber \\
&&=-g_j\left(i\,\gamma-\Omega+\omega_{j}\right)e^{-i\omega_j t},
\end{eqnarray}
with $a=\gamma$, $b=\omega_0+\Omega$, $c=\Gamma\gamma/2-\omega_0\Omega$, and $d=\gamma\omega_0$. These equations represent second-order complex coefficient linear homogeneous (Eq.~(\ref{eq141})) or inhomogeneous (Eq.~(\ref{eq232})) differential equations, the solutions of which are provided in the appendix.

It is noteworthy that this approach is applicable across a broad range of temperatures. The system is assumed to possess an initial state $\rho_0 \bigotimes \Pi_j \rho_j$, with the system and reservoir being in thermal equilibrium, expressed as $\rho_j=\sum_{n_j}\lvert n_j\rangle \langle n_j| e^{-\beta_j n_j \hbar \omega_j}/Z$. Here, $Z=\sum_{n_j=0}^{\infty}e^{-\beta_j n_j\hbar \omega_j}$ is the partition function, where $j=0$ corresponds to the system. $\beta_j=1/(k_BT_j)$ are defined with $T_0\equiv T_s$ the system temperature, $T_j\equiv T_b\,(\forall j\neq 0)$ is the temperature of the $j$th oscillator, and $k_B$ is the Boltzmann constant. We will take $k_B=1.0$ and define $\beta=1/T_0$. The average value $\left\langle a^{\dagger}(t)a(t) \right\rangle$ for the AEN is given by
\begin{eqnarray}
&&\left\langle a^{\dagger}(t)a(t) \right\rangle\nonumber\\
&=&\left|A(t) \right|^2 \left\langle a^{\dagger}(0)a(0) \right\rangle+\sum_{j} \left|B_j(t) \right|^2 \left\langle b_j^{\dagger}(0)b_j(0) \right\rangle \notag \\
&=&\left|A(t) \right|^2 \frac{1}{e^{\beta\omega_0}-1}+\sum_{j} \left|B_j(t) \right|^2 \frac{1}{e^{\beta_j\omega_j}-1}.
\end{eqnarray}
With a large number of very close modes, i.e., the continuous limit, the summation can be transformed into an integral. Then, the second term can be rewritten as
\begin{eqnarray}
	\sum_{j} \left|B_j(t) \right|^2 \frac{1}{e^{\beta_j\omega_j}-1}
	=\int
	d\omega^{\prime}J(\omega^{\prime})f(\omega^{\prime}) \frac{1}{e^{\beta^{\prime}\omega^{\prime}}-1}.\nonumber\\
	\label{eq43}
\end{eqnarray}
Upon solving for $A(t)$ and $B_j(t)$, the dynamics of the average value can be determined.

\section{Results and Discussions}

Let us introduce the real parameter $h>0$ and define $\omega_0=h\Omega$.  First, we explore the scenario where the central frequency is set as $\Omega=\omega_0$ ($h=1$). Note that the parameter $\Omega$, determines the oscillation speed of the correlation function \cite{Xinyu}. Notably, in non-Markovian environments, a back reaction emerges, similar to an external driving force acting on the system. The resonance condition, when the frequency of the external driving force matches the intrinsic frequency of the system, leads to a peak in the amplitude of the system oscillator. For $\Omega=\omega_0$, the parameters are given by $b=2\omega_0$, $d=ab/2$.

For the resonance case, we examine three scenarios:

(1) Strong Markovian regime $(\gamma>2\Gamma)$, $\Delta_2\equiv \gamma (2\Gamma-\gamma)<0$. The solution of Eqs.~(\ref{eq141}) and (\ref{eq232}) can be given analytically,
\begin{eqnarray}
A(t)=C_1e^{(\frac{-\gamma+\sqrt{-\Delta_2}}{2}- i\omega_0)t}+C_2e^{-(\frac{\gamma+\sqrt{-\Delta_2}}{2}+i\omega_0)t},
\end{eqnarray}
with $C_1=\frac{\gamma+\sqrt{-\Delta_2}}{2\sqrt{-\Delta_2}}$ and $C_2=1-C_1=\frac{\sqrt{-\Delta_2}-\gamma}{2\sqrt{-\Delta_2}}$ ,
and
\begin{widetext}
\begin{eqnarray}
B_j(t)=g_j[C_1^{j}e^{(\frac{-\gamma+\sqrt{-\Delta_2}}{2}- i\omega_0)t}+C_2^{j}e^{-(\frac{\gamma+\sqrt{-\Delta_2}}{2}+i\omega_0)t}+C_3^je^{-i\omega_j t}],
\end{eqnarray}
with $C_1^j=\frac{-[\frac{\gamma+\sqrt{-\Delta_2}}{2}+i(\omega_0-\omega_j)]C_3^j-i}{\sqrt{-\Delta_2}}$,
$C_3^j=-\frac{\omega_{j}-\omega_{0}+i,\gamma}{\Phi(-i\omega_j)}$, and $C_2^j=-C_3^j-C_1^j$. Here, $\Phi(-i\omega_j)=i\gamma(\omega_0-\omega_j)-(\omega_0-\omega_j)^2+\frac{\Gamma\gamma}{2}\neq0$ has been used.
Thus, we have $f(\omega')$ defined in Eq.~(\ref{eq43})
\begin{eqnarray}
&&f(\omega^{\prime})=[(\left|C_1^j \right|^2e^{(-\gamma+\sqrt{-\Delta_2}) t}+\left|C_2^j \right|^2e^{(-\gamma-\sqrt{-\Delta_2}) t}) +\left|C_3^j \right|^2 \nonumber \\
&+&2\text{Re}(C_1^jC_2^{j*})e^{-\gamma t}+2[\text{Re}(C_1^jC_3^{j*})\cos(\omega_0 -\omega_j)t+\text{Im}(C_1^jC_3^{j*})\sin(\omega_0 -\omega_j)t] e^{\frac{(-\gamma+\sqrt{-\Delta_2})}{2} t} \nonumber \\
&+&2[\text{Re}(C_2^jC_3^{j*})\cos(\omega_0 -\omega_j)t+\text{Im}(C_2^jC_3^{j*})\sin(\omega_0 -\omega_j)t] e^{\frac{(-\gamma-\sqrt{-\Delta_2})}{2} t}].
\label{eq49}
\end{eqnarray}

(2) The critical point $(\gamma=2\Gamma)$, $\Delta_2=0$. The solutions are
\begin{eqnarray}
	A(t)=\left(\frac{\gamma}{2}t+1\right)e^{-(\frac{\gamma}{2}+i\omega_0)t},
\end{eqnarray}
and
\begin{eqnarray}
	B_j(t)=(C_1^jt+C_2^j)g_je^{-(\frac{\gamma}{2}+i\omega_0)t}+g_jC_3^je^{-i\omega_j t},
\end{eqnarray}
with
\begin{eqnarray}
C_1^j=-i-\left(\Gamma+i\,\left(\omega_{0}-\omega_{j}\right)\right)\,C_{3},
C_2^j=-C_3^j.
\end{eqnarray}
Then, we have
\begin{eqnarray}
	&&f(\omega^{\prime})=\left| (C_1^jt+C_2^j) \right|^2 e^{-\gamma t}+\left| C_3^j \right|^2+2[\text{Re}((C_1^{\prime*}t+C_2^{\prime*})C_3^j)\cos(\omega_0-\omega^{\prime})t+\text{Im}((C_1^{\prime*}t+C_2^{\prime*})C_3^j)\sin(\omega_0-\omega^{\prime})t]e^{\frac{-\gamma }{2}t}. \nonumber
	\label{eq49}
\end{eqnarray}

(3) The strong non-Markovian regime $(\gamma<2\Gamma)$, $\Delta_2>0$. The coefficients  can be written as
\begin{eqnarray}
	A(t)=C_1e^{-(\frac{\gamma}{2}+\frac{-2\omega_0+\sqrt{\gamma(2\Gamma-\gamma)}}{2}i)t}+C_2e^{-(\frac{\gamma}{2}+\frac{2\omega_0+\sqrt{\gamma(2\Gamma-\gamma)}}{2}i)t},
\end{eqnarray}
with $C_1=\frac{\gamma+2\omega_0+(\sqrt{\gamma(2\Gamma-\gamma)}-2\omega_0)i}{4\omega_0}$, $C_2=\frac{2\omega_0-\gamma-(\sqrt{\gamma(2\Gamma-\gamma)}-2\omega_0)i}{4\omega_0}$, and
\begin{eqnarray} B_j(t)=g_j[C_1^{j}e^{-(\frac{\gamma}{2}+\frac{-2\omega_0+\sqrt{\gamma(2\Gamma-\gamma)}}{2}i)t}+C_2^{j}e^{-(\frac{\gamma}{2}+\frac{2\omega_0+\sqrt{\gamma(2\Gamma-\gamma)}}{2}i)t}+C_3^je^{-i\omega_j t}],\notag
\end{eqnarray}
with $C_1^j=\frac{1}{2\omega_0}[\frac{(-\frac{2\omega_0+\sqrt{\gamma(2\Gamma-\gamma)}}{2}+\frac{\gamma}{2}i)\omega_j+\omega_j^2}{\Phi(-i\omega_j)}+1]$,  $C_3^j=-\frac{\omega_{j}-\omega_{0}+i\,\gamma}{\Phi(-i\omega_j)}$,$C_2^j=-C_3^j-C_1^j$.
Then, we obtain
\begin{eqnarray}
	&&f(\omega^{\prime})=\bigg[\left(\left|C_1^j \right|^2+\left|C_2^j \right|^2\right) e^{-\gamma t}+\left|C_3^j \right|^2 \nonumber \\
	&+&2[\text{Re}(C_1^jC_2^{j*})\cos 2\omega_0 t-\text{Im}(C_1^jC_2^{j*})\sin 2\omega_0 t]e^{-\gamma t}+2[Re(C_1^jC_3^{j*})\cos(\frac{\sqrt{\Delta_2}}{2}\nonumber \\
     &-&\omega_0 +\omega^{\prime})t-\text{Im}(C_1^jC_3^{j*})sin(\frac{\sqrt{\Delta_2}}{2}-\omega_0 +\omega^{\prime})t]e^{-\frac{\gamma}{2} t} \nonumber \\
	&+&2[\text{Re}(C_2^jC_3^{j*})\cos(\frac{\sqrt{\Delta_2}}{2}+\omega_0 +\omega^{\prime})t-\text{Im}(C_2^jC_3^{j*})\sin(\frac{\sqrt{\Delta_2}}{2}+\omega +\omega^{\prime})t]e^{-\frac{\gamma}{2} t}\bigg].
	\label{eq44}
\end{eqnarray}

For all three cases mentioned above, $\left|A(t) \right|^2 \rightarrow 0$ as $t \rightarrow +\infty$. Thus all terms in $f(\omega^{\prime})$ decay to zero except $|C_3^j|^2$, leading to the steady-state AEN
\begin{eqnarray}
N&=&\sum_{j} \left|B_j(t) \right|^2 \frac{1}{e^{\beta_j\omega_j}-1}\nonumber\\
&=&\frac{\Gamma }{2\pi}\int d\omega^{\prime}\frac{\gamma^2}{(\omega^{\prime}-\omega_0)^2+\gamma^2}
\frac{\left(\left(\omega'-\omega_{0}\right)^{2}+\gamma^{2}\right)}{\left(\frac{\Gamma \gamma}{2}-\left(\omega_{0}-\omega'\right)^{2}\right)^{2}+\gamma^{2}\left(\omega'-\omega_{0}\right)^{2}}
\frac{1}{e^{\beta^{\prime}\omega^{\prime}}-1}.
\label{e11}
\end{eqnarray}
The AEN is then constant, depending solely on the bath parameters $\gamma$, $\Gamma$, and $T_b$. An interesting  observation is that the steady-state AEN is independent of the system temperature $T_s$.

When $\gamma\rightarrow 0$, the spectral density becomes $J(\omega'-\omega_0)=\Gamma\delta(\omega'-\omega_0)$, resulting in
\begin{eqnarray}
\sum_{j} &&\left|B_j(t) \right|^2 \frac{1}{e^{\beta_j\omega_j}-1}\nonumber\\
&=&\Gamma \int d\omega^{\prime}\delta(\omega'-\omega_0)
\frac{\left(\left(\omega'-\omega_{0}\right)^{2}+\gamma^{2}\right)}{\left(\frac{\Gamma \gamma}{2}-\left(\omega_{0}-\omega'\right)^{2}\right)^{2}+\gamma^{2}\left(\omega'-\omega_{0}\right)^{2}}
\frac{1}{e^{\beta^{\prime}\omega^{\prime}}-1}. \nonumber\\
&=&\frac{4\gamma}{\Gamma}
\frac{1}{e^{\beta^{\prime}\omega_0}-1}.
\label{enon}
\end{eqnarray}
\end{widetext}
In the Markovian limit, i.e.,$\gamma\rightarrow\infty$, it is easy to verify
$$J(\omega)=\lim_{\gamma\rightarrow\infty}\frac{\Gamma \gamma^2/2\pi}{(\omega-\Omega)^2+\gamma^2}=\frac{\Gamma}{2\pi},$$
and
\begin{eqnarray}
&&\lim_{\gamma\rightarrow\infty}\frac{\left(\left(\omega'-\omega_{0}\right)^{2}+\gamma^{2}\right)}
{\left(\frac{\Gamma\gamma}{2}-\left(\omega_{0}-\omega'\right)^{2}\right)^{2}+\gamma^{2}\left(\omega'-\omega_{0}\right)^{2}} \nonumber \\
&&=\frac{1}
{\Gamma^{2}/4+\left(\omega_{0}-\omega'\right)^{2}}.
\end{eqnarray}
Therefore, we can obtain the AEN in the Markovian limit
\begin{eqnarray}
\lim_{\gamma\rightarrow\infty}\left\langle a^{\dagger}(t)a(t) \right\rangle
&=&\int d\omega^{\prime}\frac{J(\omega ')}
{\Gamma^{2}/4+\left(\omega_{0}-\omega'\right)^{2}}\frac{1}{\exp(\beta'\omega')-1}\nonumber\\
&=& \frac{1}{\exp(\beta'\omega_0)-1}.
\label{Markovianlimit}
\end{eqnarray}
Clearly, in the long time and Markovian limit, $N$ only depends on the bath temperature. In other words, the system reaches thermal equilibrium with the bath, and $\Gamma$ alone determines the dissipation rate.

Interestingly, this last result can be proven in another way.  Consider $\gamma \rightarrow +\infty$, the correlation function $G(t,s)\rightarrow \frac{\Gamma}{2}\delta(t,s)$, and $J(\omega)=\frac{\Gamma}{2\pi}$. Eq.~(\ref{eq11}) becomes
\begin{eqnarray}
\dot{A(t)}=-i\omega_0 A(t)-\frac{\Gamma}{2}A(t).
\end{eqnarray}
With the initial condition $A(0)=1$, the solution is
\begin{eqnarray}
A(t)=e^{-(\frac{\Gamma}{2}+i\omega_0)t}.
\end{eqnarray}
Moreover, Eq.~(\ref{eq12}) becomes
\begin{eqnarray}
\dot{B_j(t)}=-(\frac{\Gamma}{2}+i\omega_0) B_j(t)-ig_je^{-i\omega_j t}.
\end{eqnarray}
which leads to 
\begin{eqnarray}
B_j(t)=\frac{g_j}{-(\omega_0-\omega_j)+i\frac{\Gamma}{2}}(e^{-i\omega_jt}-e^{-(i\omega_0+\frac{\Gamma}{2})t}).
\label{75}
\end{eqnarray}
Here the initial conditions $B_j(0)=0$ have been used. Then
\begin{eqnarray}
\left|A(t) \right|^2=e^{-\Gamma t}.
\end{eqnarray}
and
\begin{eqnarray}
&&\left|B_j(t) \right|^2\notag \\
&=&\frac{\left|g_j \right|^2}{(\omega_0-\omega_j)^2+(\frac{\Gamma}{2})^2}(1+e^{-\Gamma t}+2\cos[(\omega_0-\omega_j)t]e^{-\frac{\Gamma}{2} t}).\notag
\end{eqnarray}
Clearly, when $t\rightarrow +\infty$, $\left|A(t) \right|^2\rightarrow 0$.
\begin{eqnarray}
\left|B_j(t) \right|^2&\rightarrow&\frac{\left|g_j \right|^2}{(\omega_0-\omega_j)^2+(\frac{\Gamma}{2})^2}.
\end{eqnarray}
By considering $\sum_{j}\lvert g_j \rvert^2 \delta(\omega-\omega_j)\rightarrow \int_{-\infty}^{+\infty} d\omega J(\omega)$, we have
\begin{eqnarray}
\left\langle a^{\dagger}(t)a(t) \right\rangle&=&\sum_j |B_j(t)|^2\frac{1}{e^{\beta_j\omega_j}-1}\nonumber\\
&=&\int_{-\infty}^{+\infty} d\omega\frac{J(\omega')}{(\omega_0-\omega_j)^2+(\frac{\Gamma}{2})^2}\frac{1}{e^{\beta^{\prime}\omega'}-1}\nonumber\\
&=&\frac{1}{e^{\beta^{\prime}\omega_0}-1}
\end{eqnarray}
where $\int_{-\infty}^{+\infty}\frac{d\omega}{(\omega_0-\omega)^2+(\frac{\Gamma}{2})^2}=2\pi/\Gamma$ has been used. This result aligns with Eq.~(\ref{Markovianlimit}).

\begin{figure}
	(a)
	\centerline{\includegraphics[width=0.8\columnwidth]{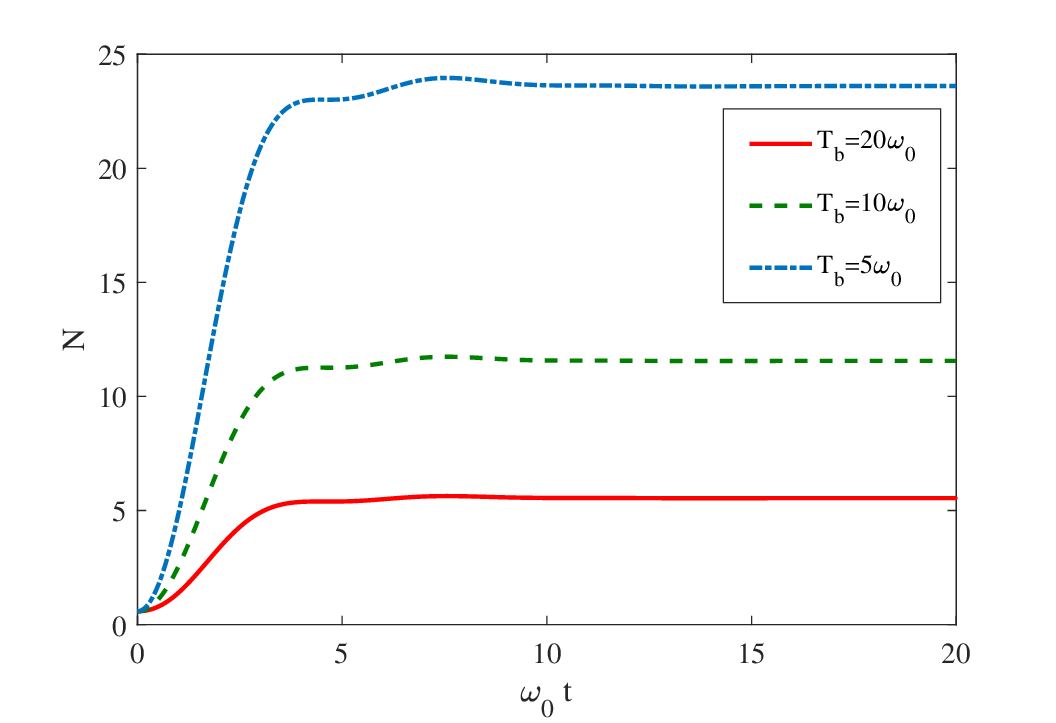}}
	(b)
	\centerline{\includegraphics[width=0.8\columnwidth]{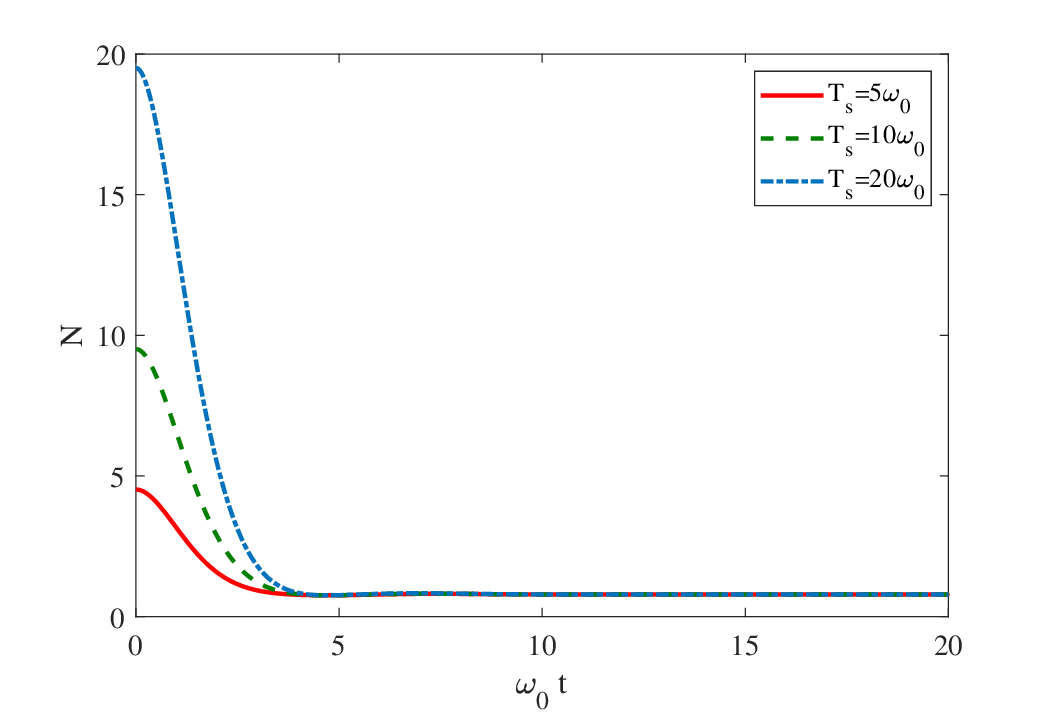}}
	(c)
	\centerline{\includegraphics[width=0.8\columnwidth]{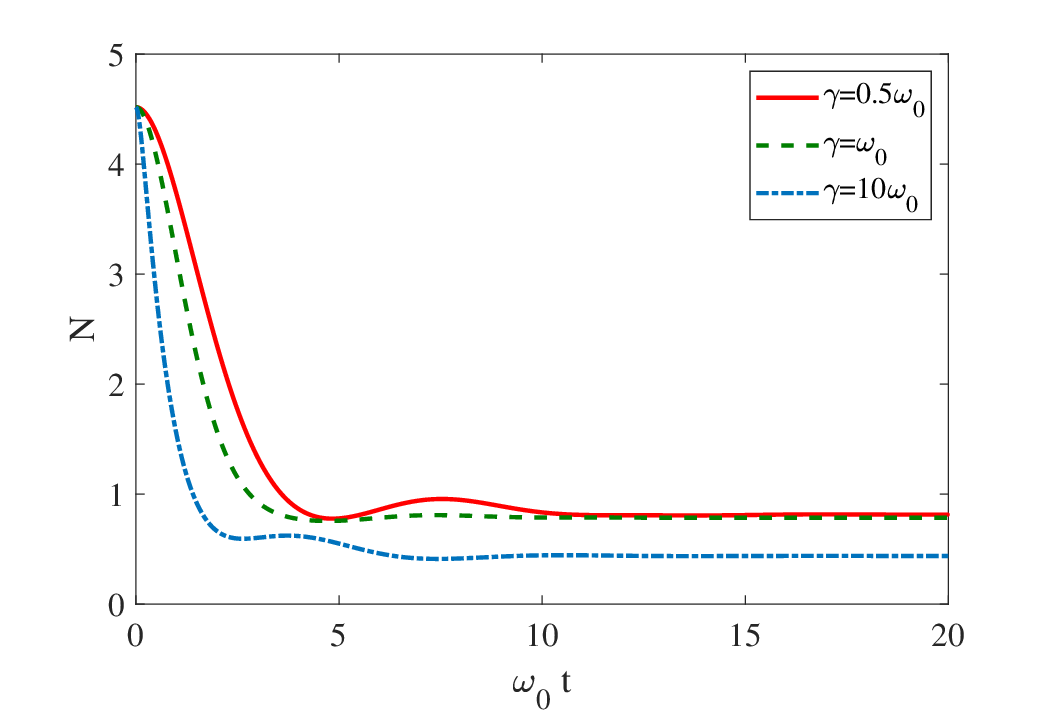}}
	(d)
	\centerline{\includegraphics[width=0.8\columnwidth]{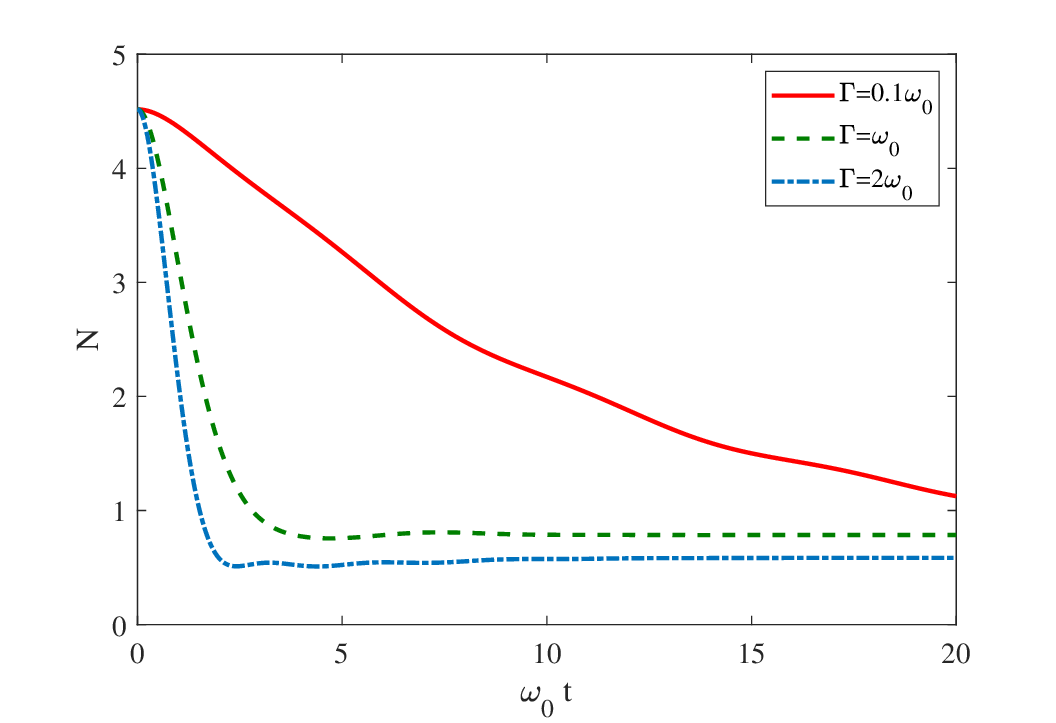}}
	
	\caption{(Color online)  $N$ versus dimensionless time $\omega_0 t$ for different (a) the bath temperature $T_b$; (b) the system temperature $T_s$; (c) the spectral width $\gamma$; (d) the dissipative rate $\Gamma $. We take the environmental temperature $T_b=\omega_0$ and set $\omega_0=\Omega=1$. $\Gamma=1$, $\gamma=1$ in (a) $T_s=10$, $\Gamma=1$, $\gamma=1$  in (b), $T_s=5$, $T_b=1$, $\Gamma=1$ in (c),  $T_s=5$, $T_b=1$, $\gamma=1$ in (d).}
	\label{Fig:1}
\end{figure}

In Fig.~\ref{Fig:1}, $N$ is plotted as a function of $\omega_0 t$.  Plots (a), (b), (c), and (d), show different bath temperatures $T_b$, system temperatures $T_s$, spectral widths $\gamma$, and dissipation rates $\Gamma$, respectively. Fig.~\ref{Fig:1}(a) illustrates that $N$ increases with rising $T_b$ for a certain $T_s$. Fig.~\ref{Fig:1}(b) shows that, for a constant $T_b$, the initial AEN $N(t=0)$ is higher with an elevated $T_s$. Moreover, $N$ converges to the same value at thermal equilibrium (long time limit), which is consistent with Eq.~(\ref{e11}). The trend continues in Fig.~\ref{Fig:1}(c), indicating that smaller $\gamma$ leads to a higher $N$. In other words, non-Markovianity contributes to an increased excitation number, even in the long time limit. As anticipated, stronger couplings $\Gamma$ correspond to smaller $N$, as depicted in Fig.~\ref{Fig:1}(d).

\begin{figure}
	(a)
	\centerline{\includegraphics[width=0.8\columnwidth]{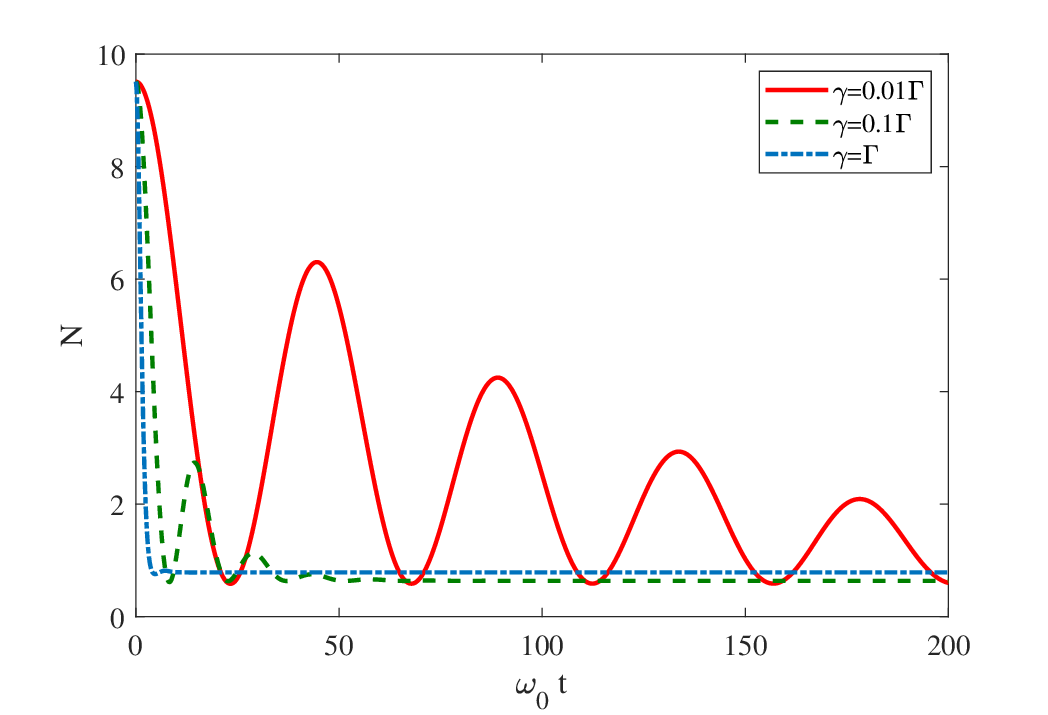}}
	(b)
	\centerline{\includegraphics[width=0.8\columnwidth]{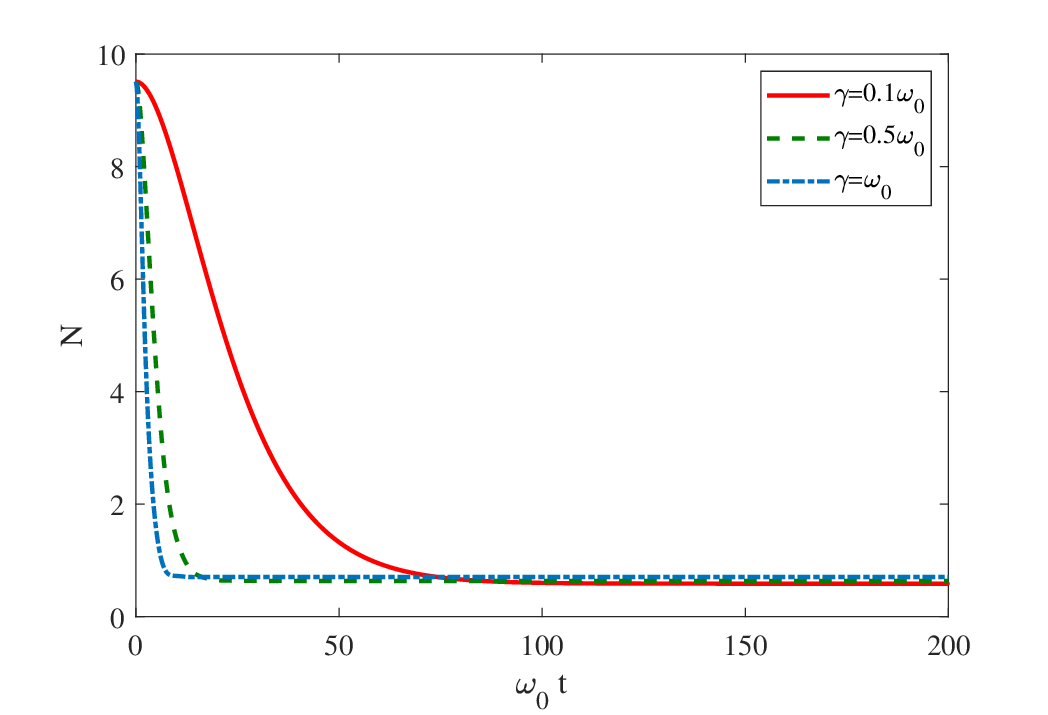}}
	
	\caption{(Color online) $N$ as a  function of dimensionless time $\omega_0 t$ for different parameter $\gamma$ (a) $\Gamma=1$, $\gamma<2\Gamma$; (b) $\gamma=2\Gamma$. For both (a) and (b), $T_s=10$, $T_b=1$, $\omega_0=\Omega=1$.}
	\label{Fig:2}
\end{figure}

The role of non-Markovianity is pivotal in the exploration of open quantum systems. Therefore, it is important to focus on the impact of the parameter $\gamma$ on system dynamics, with particular emphasis on the strong non-Markovian ($\gamma<2\Gamma$) and critical $\gamma=2\Gamma$ regions.
Fig.~\ref{Fig:2}(a) and (b) illustrates how $N$ depends on $\omega_0 t$ for these two scenarios. In Fig.~\ref{Fig:2}(a), a pronounced oscillation of $N$ with $\omega_0 t$ is observed for a strongly non-Markovian bath ($\gamma=0.01\Gamma$). This oscillation diminishes as $\gamma$ increases and completely disappears in critical cases, as depicted in Fig.~\ref{Fig:2}(b). In both instances, a non-zero steady state for $N$ is achieved. Non-Markovianity from the bath facilitates an augmentation in AEN, as shown also in Fig.~\ref{Fig:1}(c). Notably, the parameter $\gamma$ governs the steady-state AEN as is evident from Eq.~(\ref{e11}).

\begin{figure}
	\centerline{\includegraphics[width=0.8\columnwidth]{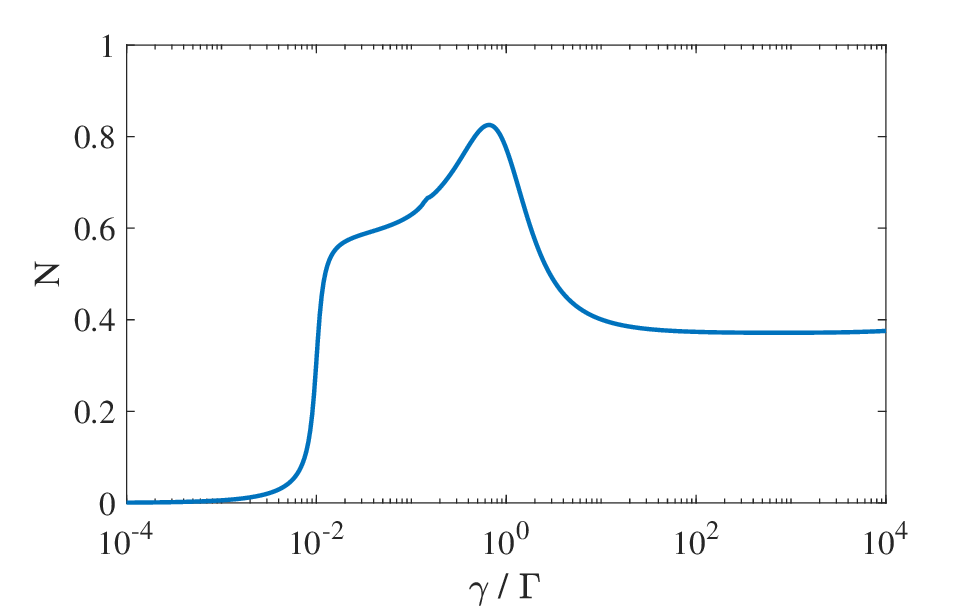}}
	\caption{(Color online) The steady-state AEN $N$ as a function of parameter $\gamma/\omega_0$   with $\Gamma=1$, $T_b=1$, $\omega_0=\Omega=1$.}
	\label{Fig:3}
\end{figure}

Fig.~\ref{Fig:3} shows the influence of non-Markovianity on the steady-state AEN, by providing the relationship between steady-state AEN and $\gamma$. The parameters are set to $\Gamma=1$ and $T_b=1$. From Fig.~\ref{Fig:3}, it is clear that steady-state AEN initially rises, reaches a peak, and subsequently decreases with increasing $\gamma$. This behavior is predicted by Eq.~(\ref{enon}). Specifically, in the limit as $\gamma\rightarrow 0$, steady-state AEN is proportional to $\gamma$, as shown in the corresponding region in Fig.~\ref{Fig:3}.
The underlying physics governing this trend lies in the fact that, as $\gamma \rightarrow 0$, only a few bath modes are coupled to the system. Consequently, a smaller steady-state AEN is observed. When $\gamma=0$, the AEN consistently oscillates between the system and the bath, resulting in a steady-state AEN of $N=0$. As $\gamma$ increases, more modes become coupled to the system, leading to a subsequent increase in $N$.

Previously the discussions have considered the system in the resonance condition. Now, the analysis is extended to the more general scenario where the system frequency $\omega_0$ deviates from the central frequency $\Omega$. In this off-resonance case ($h\neq1$), the system dynamics are influenced by the value of $h$ and the other parameters take the following values: $a=\gamma, b=(h+1)\Omega, c=\Gamma\gamma/2-h\Omega^2, d=\gamma h \Omega$, and $\Delta_1=(1-h)\gamma \Omega$.

Specifically, consider the condition where $c$ does not equal $(a^2-b^2)/4$, i.e., $\Gamma\gamma/2-h\Omega^2\neq \frac{\gamma^2-((h+1)\Omega)^2}{4}$.

\begin{widetext}
(1) When $\Delta_1<0,(h>1) $,  the solution of Eq.~(\ref{eq141}) will be Eq.~(\ref{eq15}) with
\begin{eqnarray} C_1=\frac{\frac{a}{2}-\frac{\sqrt{-\Delta_1}}{2}\sqrt{\sqrt{\lambda_1^2+1}-\lambda_1}+\frac{1}{2}\bigg(b-2h\Omega+\frac{\sqrt{-\Delta_1}}{\sqrt{\sqrt{\lambda_1^2+1}-\lambda_1}}\bigg)i}
{-\sqrt{-\Delta_1}\sqrt{\sqrt{\lambda_1^2+1}-\lambda_1}+\frac{\sqrt{-\Delta_1}}{\sqrt{\sqrt{\lambda_1^2+1}-\lambda_1}}i},
\end{eqnarray}
and
\begin{eqnarray}
	C_2=1-C_1.
\end{eqnarray}
 Here $\lam_1=\frac{2\Gamma-\gamma}{2(1-h)\Omega}+\frac{(1-h)\Omega}{2\gamma}$ has been used.

The solution of Eqs.~(\ref{eq232}) is
\begin{eqnarray} B_j(t)=g_j\left[C_1^{j}e^{\big[-\frac{a}{2}-\frac{\sqrt{-\Delta_1}}{2}\sqrt{\sqrt{\lambda_1^2+1}-\lambda_1}-\frac{1}{2}\big(b-\frac{\sqrt{-\Delta_1}}{\sqrt{\sqrt{\lambda_1^2+1}-\lambda_1}}\big)i\big]t}
\!\!\! +C_2^{j}e^{\big[-\frac{a}{2}+\frac{\sqrt{-\Delta_1}}{2}\sqrt{\sqrt{\lambda_1^2+1}-\lambda_1}-\frac{1}{2}\big(b+\frac{\sqrt{-\Delta_1}}{\sqrt{\sqrt{\lambda_1^2+1}-\lambda_1}}\big)i\big]t}+
C_3^je^{-i\omega_j t}\right],\notag
\end{eqnarray}
where
\begin{eqnarray}
C_1^j=-\frac{\big[-\frac{a}{2}+\frac{\sqrt{-\Delta_1}}{2}\sqrt{\sqrt{\lambda_1^2+1}-\lambda_1}-\frac{1}{2}\big(b+\frac{\sqrt{-\Delta_1}}{\sqrt{\sqrt{\lambda_1^2+1}-\lambda_1}}\big)i\big]
C_3^j-i\omega_jC_3^j+ig_j}{-\sqrt{-\Delta_1}\sqrt{\sqrt{\lambda_1^2+1}-\lambda_1}+\frac{\sqrt{-\Delta_1}}
{\sqrt{\sqrt{\lambda_1^2+1}-\lambda_1}}i},
\end{eqnarray}
and
\begin{eqnarray}
	C_2^j=C_3^j-C_1^j,
\end{eqnarray}
with $C_3^j=-\frac{\omega_{j}-\Omega+i\,\gamma}{\Phi(-i\omega_j)}.$

(2) When $\Delta_1>0,(h<1) $, the solution of Eq.~(\ref{eq232}) will be Eq.~(\ref{eq15}) with
\begin{eqnarray}
C_1=\frac{-i\omega_0+\big[\frac{a}{2}+\frac{\sqrt{\Delta_1}}{2}\sqrt{\sqrt{\lambda_1^2+1}-\lambda_1}+\frac{1}{2}\big(b+\frac{\sqrt{\Delta_1}}{\sqrt{\sqrt{\lambda_1^2+1}
-\lambda_1}}\big)i\big]}{\sqrt{\Delta_1}\sqrt{\sqrt{\lambda_1^2+1}-\lambda_1}+i\frac{\sqrt{\Delta_1}}{\sqrt{\sqrt{\lambda_1^2+1}-\lambda_1}} },
\end{eqnarray}
\begin{eqnarray}
	C_2=1-C_1.
\end{eqnarray}
The solution of Eqs.~(\ref{eq232}) reads
\begin{eqnarray}
	B_j(t)&=&g_j\left[C_1^{j}e^{\big[-\frac{a}{2}-\frac{\sqrt{-\Delta_1}}{2}\sqrt{\sqrt{\lambda_1^2+1}-\lambda_1}-\frac{1}{2}\big(b-\frac{\sqrt{-\Delta_1}}{\sqrt{\sqrt{\lambda_1^2+1}-\lambda_1}}\big)i]t}
\!\!\! +C_2^{j}e^{\big[-\frac{a}{2}+\frac{\sqrt{-\Delta_1}}{2}\sqrt{\sqrt{\lambda_1^2+1}-\lambda_1}-\frac{1}{2}\big(b+\frac{\sqrt{-\Delta_1}}{\sqrt{\sqrt{\lambda_1^2+1}-\lambda_1}}\big)i\big]t}
+C_3^je^{-i\omega_j t}\right], \notag
\end{eqnarray}
where
\begin{eqnarray} C_1^j=\frac{-ig_j+\omega_jC_3^j+C_3^j\big[-\frac{a}{2}-\frac{\sqrt{\Delta_1}}{2}\sqrt{\sqrt{\lambda_1^2+1}-\lambda_1}-\frac{1}{2}\big(b
+\frac{\sqrt{\Delta_1}}{\sqrt{\sqrt{\lambda_1^2+1}-\lambda_1}}\big)i\big]}{\sqrt{\Delta_1}\sqrt{\sqrt{\lambda_1^2+1}-\lambda_1}
+i\frac{\sqrt{\Delta_1}}{\sqrt{\sqrt{\lambda_1^2+1}-\lambda_1}}},
\end{eqnarray}
\begin{eqnarray}
	C_2^j=-C_3^j-C_1^j,
\end{eqnarray}
with
\begin{eqnarray}
C_3^j=-\frac{\omega_{j}-\Omega+i\,\gamma}{\Phi(-i\omega_j)},
\end{eqnarray}
where $\lam_1=\frac{2\Gamma\gamma-4h\Omega^2+((h+1)\Omega)^2-\gamma^2}{2\Delta_1}$ and
$\Phi(-i\omega_j)=i\gamma(h\Omega-\omega_j)-(h\Omega-\omega_j)^2+\frac{\Gamma\gamma}{2}\neq0$.
\end{widetext}

\begin{figure}
	(a)
	\centerline{\includegraphics[width=0.8\columnwidth]{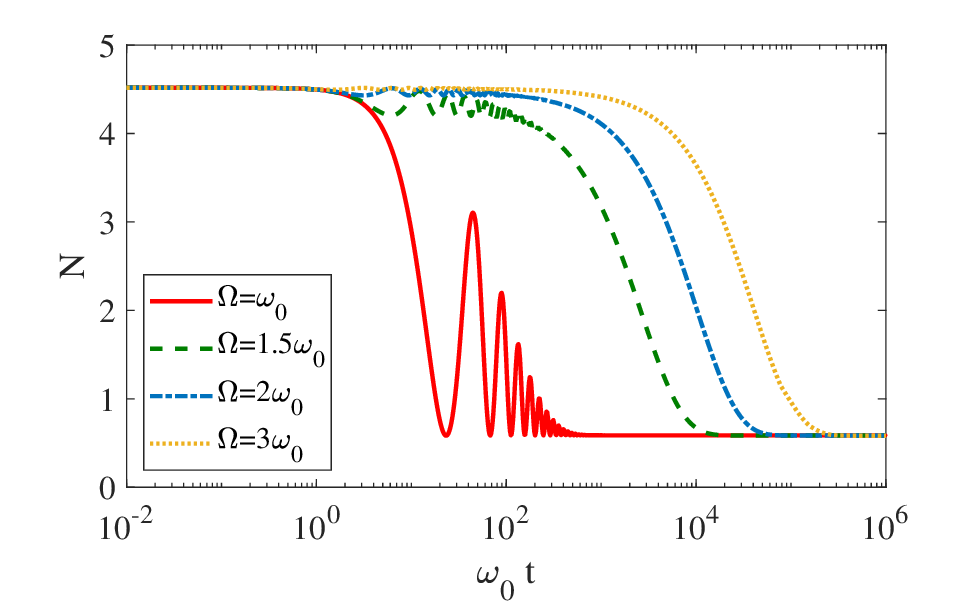}}
       (b)
       \centerline{\includegraphics[width=0.8\columnwidth]{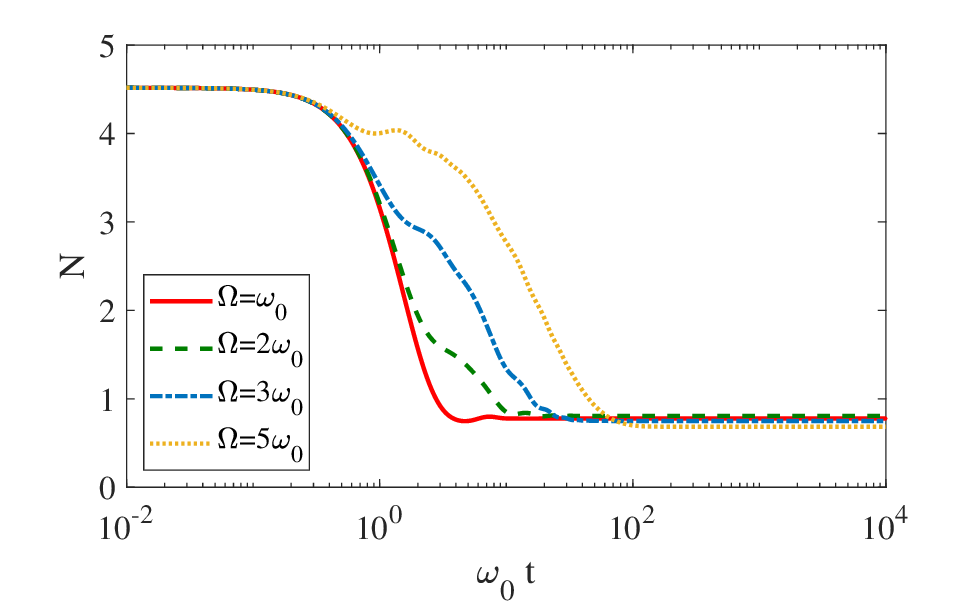}}
	  (c)
       \centerline{\includegraphics[width=0.8\columnwidth]{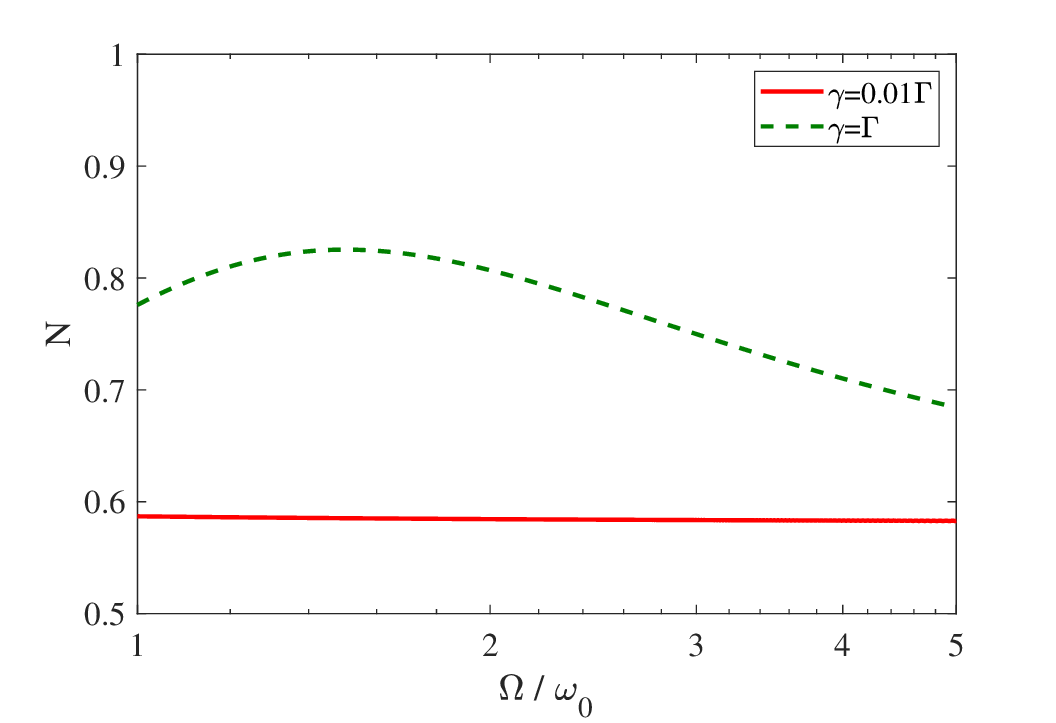}}
	\caption{ (Color online) The AEN as a function of the dimensionless time $\omega_0 t$ with different $\Omega$ for (a) $\gamma=0.01\Gamma$ and (b) $\gamma=\Gamma$. (c) The AEN in the long-time limit as a function of $\Omega/\omega_0$. The other parameters are set to be $T_b=\omega_0$, $T_s=5\omega_0$, and $\Gamma=\omega_0$. We set $\omega_0=1$ as a unit. }
	\label{Fig:5}
\end{figure}

In Fig.~\ref{Fig:5}(a) the AEN is plotted as a function of $\omega_0 t$ for the strong
non-Markovian case with $\Gamma=0.01\gamma$ and in Fig.~\ref{Fig:5}(b) the weak non-Markovian
case with $\gamma=\Gamma$. Notably, as the central frequency $\Omega/\omega_0$ increases, the characteristic time for dissipation into the steady state extends in both strong and weak non-Markovian
dynamics. Particularly, the harmonic oscillator retains its initial energy much longer in the strong
non-Markovian case compared to the weak non-Markovian case. Furthermore, the steady-state AENs are found to depend on $\Omega$ for $\gamma=\Gamma$, whereas this dependence
is less evident in the case with $\gamma=0.01\Gamma$. To emphasize this difference, the steady-state
AENs are explicitly depicted in Fig.~\ref{Fig:5}(c). Notably, in the strong non-Markovian
case ($\gamma/\Gamma\ll 1$), the steady-state AENs exhibit insensitivity to the central
frequency of the reservoir. However, with an increase in $\gamma/\Gamma$, the center frequency $\Omega$
starts influencing the AENs at the steady state in a non-monotonic manner.

\begin{figure}
	(a)
	\centerline{\includegraphics[width=0.8\columnwidth]{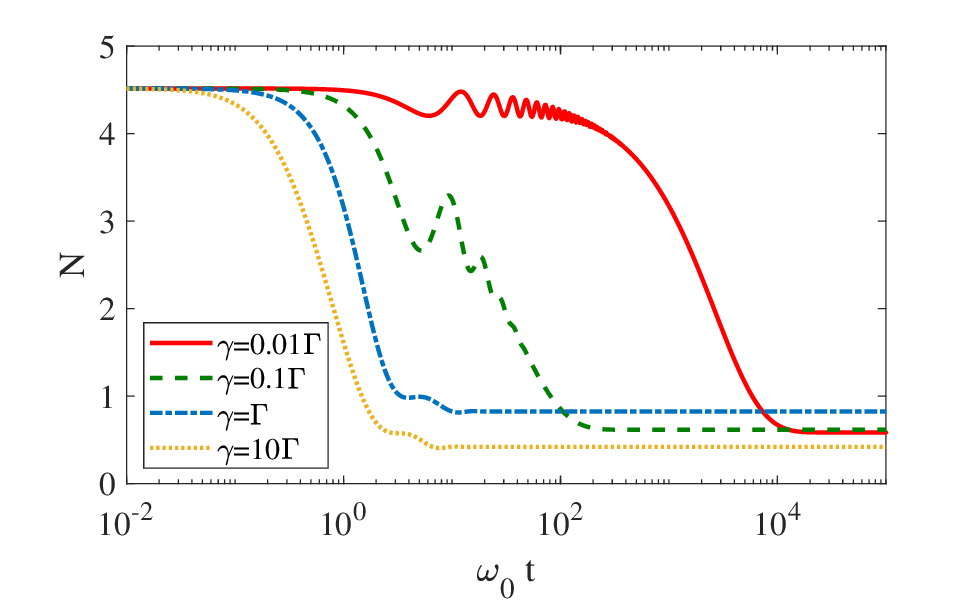}}
       (b)
       \centerline{\includegraphics[width=0.8\columnwidth]{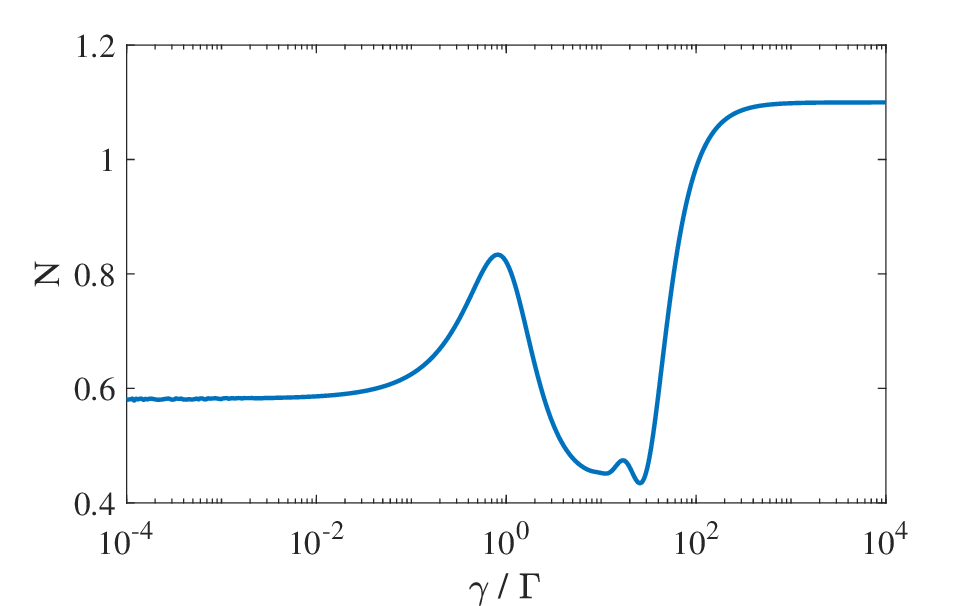}}
	\caption{ (Color online) The AEN as a function of the dimensionless time $\omega_0 t$ with different $\gamma/\Gamma$ for $\Omega=1.2\omega_0$. (c) The AEN in the long-time limit as a function of $\gamma/\Gamma_0$. The other parameters are set to be $T_b=\omega_0$, $T_s=5\omega_0$, $\Omega=1.2\omega_0$. We set $\omega_0=1$ as a unit. }
	\label{Fig:6}
\end{figure}

To show how $N$ varies with time, Fig.~\ref{Fig:6}(a) shows the relation.
It is evident that the characteristic time of non-Markovian dissipation is significantly greater than that of the Markovian
dynamics. In the case of $\gamma=0$, the dynamics can be conceptualized as a harmonic oscillator coupling
to a single mode with a detuning $\omega_0-\Omega$. Additionally, distinct steady-state AENs
are observed for different values of $\gamma/\Gamma$. This variation in steady-state AENs is illustrated in Fig.~\ref{Fig:6}(b).
Examining the figure closely, it can be seen that as $\gamma/\Gamma\rightarrow 0$, the steady-state AEN
converges to a constant, representing the resident energy that remains unaffected by exchanges with the
single-mode reservoir. Conversely, for $\gamma/\Gamma\rightarrow \infty$, the steady-state AEN approaches its Markovian limit. A noteworthy observation is that, in comparison with the
resonant case (Fig.~\ref{Fig:3}(b)), the off-resonant scenario yields a higher steady-state AEN.

\begin{figure}
	(a)
	\centerline{\includegraphics[width=0.8\columnwidth]{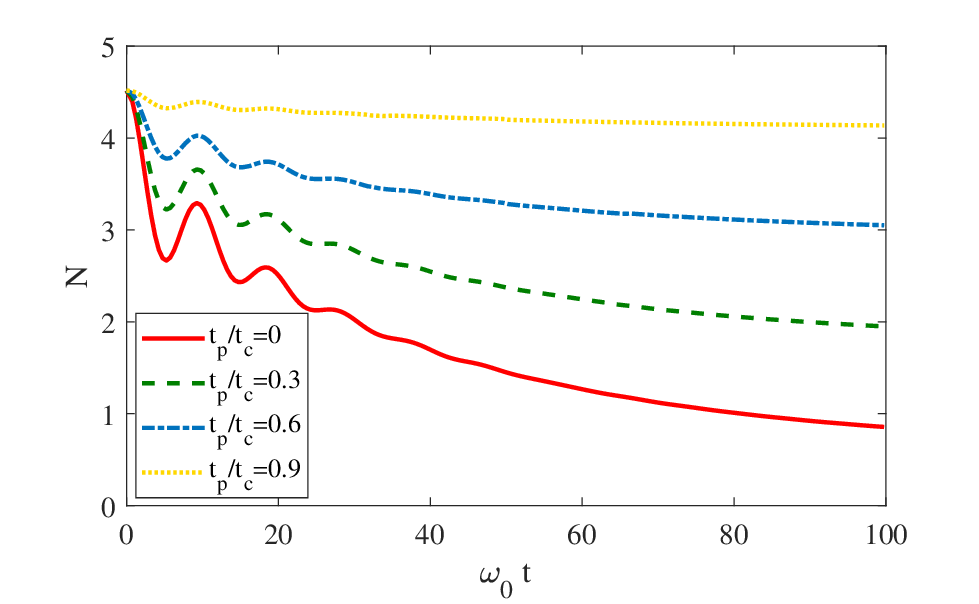}}
    (b)
       \centerline{\includegraphics[width=0.8\columnwidth]{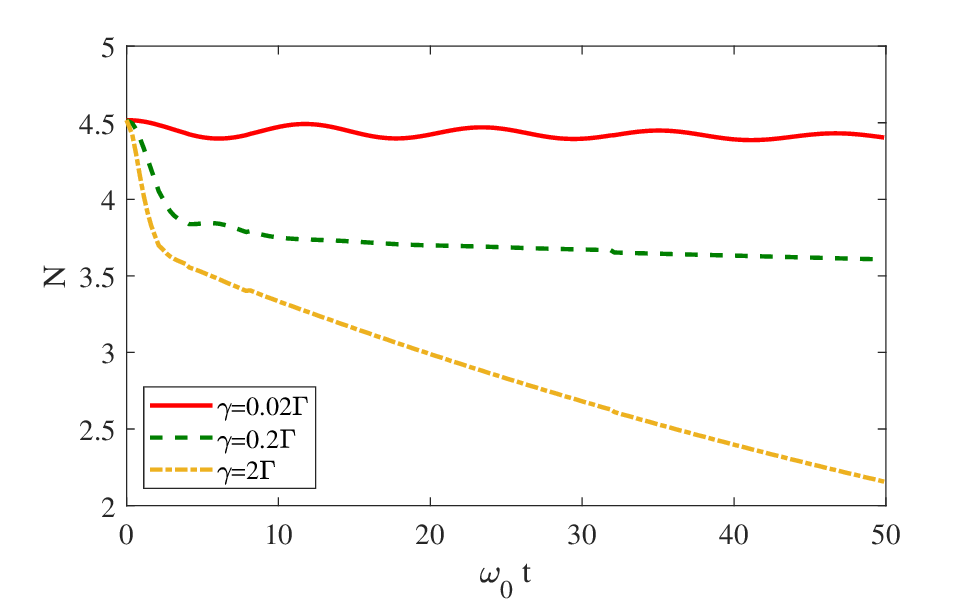}}
	  (c)
       \centerline{\includegraphics[width=0.8\columnwidth]{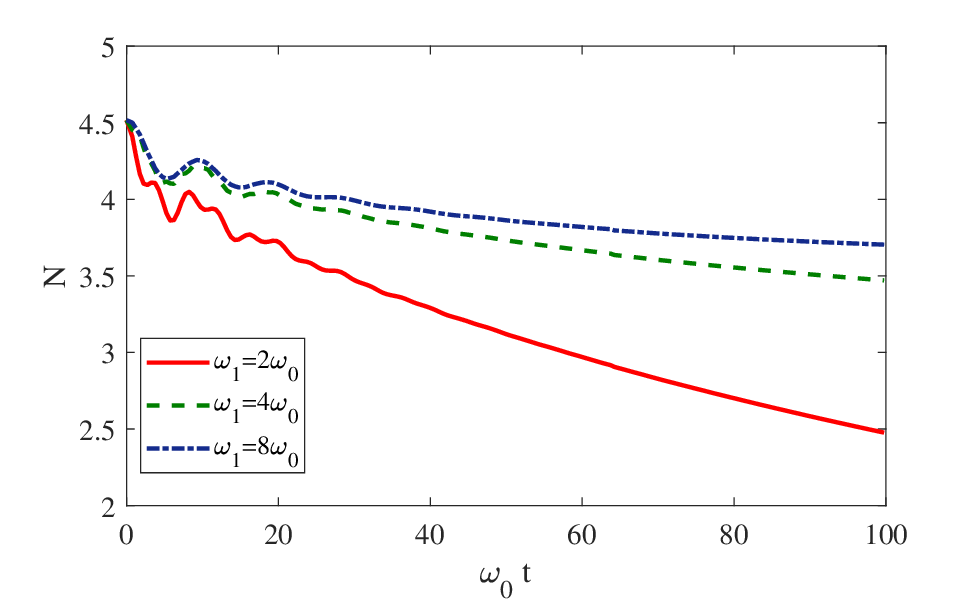}}
	\caption{(Color online) The evolution of the AEN  with different $\gamma/\Gamma$ for
(a) different ratios of pulse duration and period $t_p/t_c$ with $\omega_1=8\omega_0$ and
$\gamma=0.1\Gamma$;  (b) different pulse amplitude $\omega_1$ with $t_p/t_c=0.6$ and
$\gamma=0.1\Gamma$; (c) for different $\gamma/\Gamma_0$ with $\omega_1=8\omega_0$ and
$t_p/t_c=0.8$. The other parameters are set to be $T_b=\omega_0$, $T_s=5\omega_0$,
$\Omega=1.2\omega_0$. We set $\omega_0=1$ as a unit. }
	\label{Fig:8}
\end{figure}

\section{The leakage elimination operator control}

The effects of decoherence induced by the environment can be significantly reduced by employing a time-dependent leakage elimination operator (LEO) \cite{Wu2002, Wu2005}, denoted as $H_{LEO}
= [C(t) -\omega_0]a^\dagger a$. This type of control pulse allows the reduction of noise by cancellation of interaction terms by acting only on the system.  Here, $C(t)$ represents the function governing external pulses to be used as LEOs on the system of interest. Specifically, consider the control field as a series of
standard rectangular pulses applied to the system, with the condition $C(t)>\omega_0$. These
pulses are defined by $C(t ) = \omega_1$ for $nt_c < t \leq n t_c +t_p$ ($n \geq 0$), and otherwise
$C(t ) = \omega_0$. Here, $\omega_1$ signifies the pulse strength, $t_c$ denotes the time period, and $t_p$ represents the pulse duration or width.

In comparison to the scenario without quantum control, represented by the red solid line in Fig.~\ref{Fig:8}(a), the application of consecutive rectangular pulses demonstrates a significant suppression of decoherence in a non-Markovian environment. Generally,
achieving decoherence elimination is more feasible with a smaller $\gamma$. This phenomenon can be understood as follows: the single-mode harmonic oscillator is coupled to a non-Markovian
Ornstein-Uhlenbeck process, analogous to the harmonic oscillator resonantly coupling to a single-mode
reservoir undergoing Markovian damping. The linewidth of the single-mode reservoir is proportionate
to $\gamma$. For small $\gamma$, the trap frequency of the harmonic oscillator undergoes effective modification, leading to an off-resonance state between the harmonic oscillator and the single-mode
reservoir when subjected to a series of rapid pulses. This off-resonance condition results in a substantial
suppression of harmonic oscillator decoherence. This is confirmed and depicted by the
solid (red) lines in Fig.~\ref{Fig:6}(a) and Fig.~\ref{Fig:8}(b) accordingly. Conversely, for larger $\gamma$,
indicative of a broader reservoir linewidth, achieving off-resonance between the harmonic oscillator and
the reservoir becomes more challenging. In other words, suppressing decoherence becomes more difficult
with larger $\gamma$. However, as demonstrated in Fig.~\ref{Fig:8}(b), decoherence can be markedly reduced
when $\gamma=0.02\Gamma$. With increasing $\gamma$, although achieving perfect decoherence suppression becomes more challenging, a notable reduction in decoherence is still observed.

The efficacy of quantum control can be significantly enhanced by adjusting both $t_c/t_p$ and
$\omega_1/\omega_0$, as illustrated in Figs. \ref{Fig:8}(a) and (c), where $N$ remains nearly constant,
indicating successful control over the desired system. In Fig.~\ref{Fig:8}(a), the ratio between the pulse duration and period, $t_c/t_p$, is shown.  This shows the expected decrease in coherent amplitude with a decrease in $t_c/t_p$. Conversely, in Figs. \ref{Fig:8}(c), the evolution of $N$ for various
pulse strengths $\omega_1$ is shown. The numerical results demonstrate that a stronger $\omega_1$ leads to a
larger final AEN.

\begin{figure}
	(a)
	\centerline{\includegraphics[width=0.8\columnwidth]{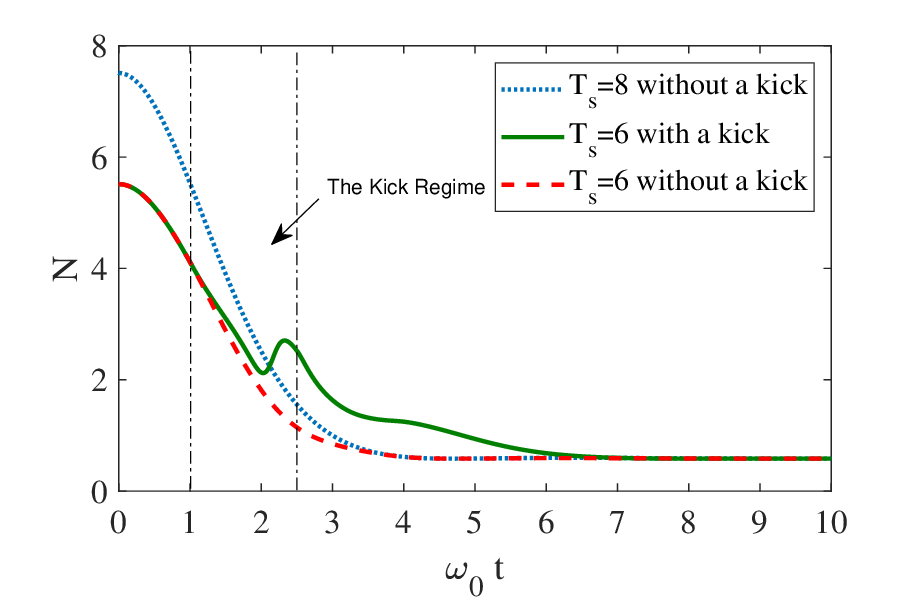}}
    (b)
       \centerline{\includegraphics[width=0.8\columnwidth]{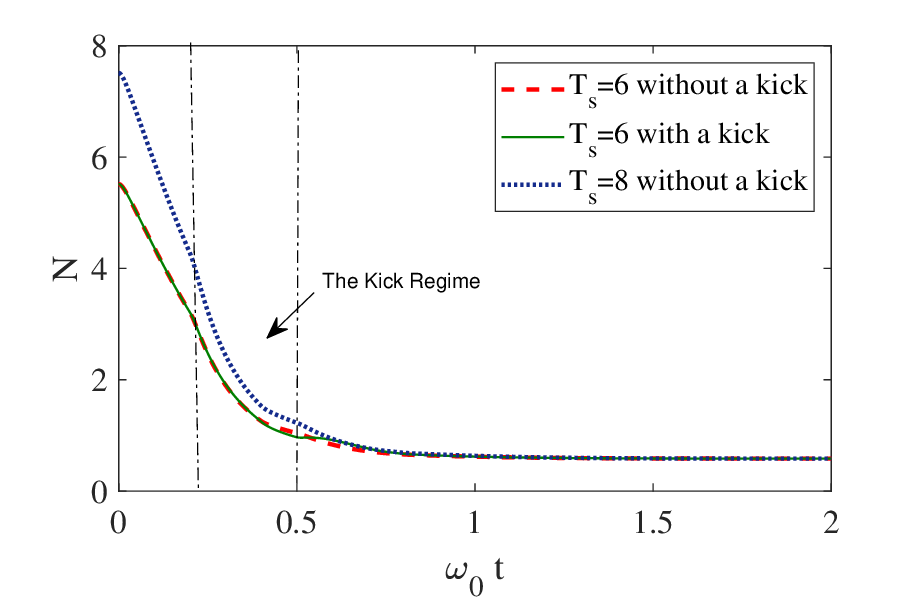}}
           (c)
       \centerline{\includegraphics[width=0.8\columnwidth]{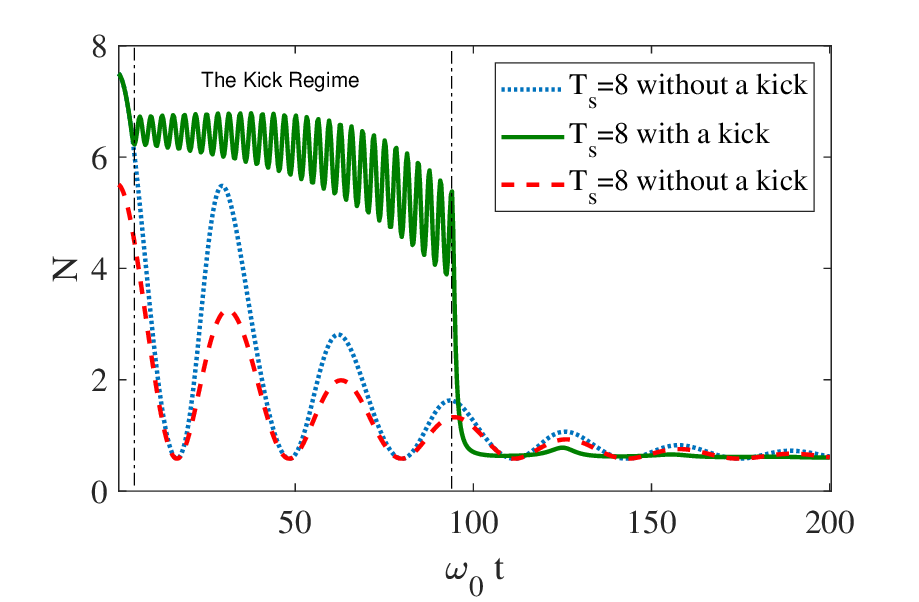}}
	\caption{(Color online) The evolution of the AEN  with different system
initial temperatures and kick pulse protocols for  (a) $\gamma=\Gamma$ with $\omega_k=5\omega_0$
 and $\Gamma=\omega_0$; (b) $\gamma=100\Gamma$ with$\omega_k=5\omega_0$
 and $\Gamma=5\omega_0$;, (c) $\gamma=0.01\Gamma$ with$\omega_k=3\omega_0$
 and $\Gamma=\omega_0$.  The other parameters are set to be $T_b=\omega_0$,   and
$\Omega=\omega_0=1$. }
	\label{Fig:9}
\end{figure}

\section{The Mpemba Effect Induced by a Kick Pulse}

The Mpemba effect occurs when two samples of identical substances, sharing all  macroscopic properties except initial temperature, exhibit a phenomenon where the initially higher-temperature
sample relaxes faster compared the lower-temperature sample. In this study, the Mpemba effect can be shown to be induced through the application of a ``kick pulse".  Specifically, this is shown here using a rectangular pulse, also referred to as a ``kick pulse", applied to an open system harmonic oscillator system.  This is observed by using a
pulse width of $\Delta t$ and a frequency of $\omega_k$. The pulse is initiated at $t_0$ and ends at $t_0 + \Delta t$.

First, non-Markovian dynamics is explored with $\gamma$ approximately equal to
$\Gamma$. In Fig.~\ref{Fig:9}(a), the evolution of the AEN is presented for
$\gamma=\Gamma$.  The numerical results for cases without the application of a kick pulse are shown with a blue dotted line for $T_s=8\omega_0$ and a red dashed line for $T_s=6\omega_0$.  In the absence of a kick pulse, the dynamics follow a normal trajectory, and no Mpemba effect is evident. However, when a kick pulse is applied to the harmonic
oscillator with a low initial system temperature, the evolution no longer exhibits monotonic decrease. This non-monotonic behavior occurs during and after the kick pulse, as illustrated by the green solid line. Consequently, the initially high-temperature oscillator
relaxes faster than the oscillator with a low initial temperature. This phenomenon can be attributed to the change of the potential boundary of the harmonic oscillator by the kick pulse. After the kick pulse, the boundary needs to return to its original state, resulting in a
slow relaxation of the oscillator system. In the Markovian limit, this effect is minimal, as shown in Fig.~\ref{Fig:9} (b). When a similar kick pulse is applied to the resonator system,
the evolution only shows minor deviations from the pulse-free case, and no Mpemba effect is observed.

Second, the case of strong non-Markovian dynamics is considered as depicted in Fig.~\ref{Fig:9}(c). The kick pulse is applied to the harmonic oscillator with a high initial temperature. The kick pulse is initiated at $t=0.02t_f$ and ended when the decay strength becomes negative. Surprisingly, after the pulse ceases, the AEN decays rapidly into a steady state with a strong decay rate and subsequently stabilizes with minor oscillations exhibiting an almost supercooling-like effect. When compared with the dynamics for the low initial temperature (red dashed line), the system with the high initial temperature undergoes quicker decay into the steady state, demonstrating the Mpemba effect induced by the kick pulse.

\section{Conclusions}

Here, the Heisenberg equation was solved exactly for an open system that consists of one oscillator interacting with a set of oscillators in which many interesting phenomena can be observed.  To exhibit these, a Lorentz spectrum is used to derive a pair of second-order complex coefficient linear differential equations governing the dynamics. The calculation of the AEN for the system reveals complex behavior in different parameter regions.  This was done for the system for two cases, on resonance and off resonance. In both scenarios, the non-Markovian effects from the baths play a crucial role in the dynamics, with the steady-state AEN being dependent on the the non-Markovian parameter $\gamma$. Notably, the number exhibits an initial increase followed by a decrease with growing $\gamma$.
The presence of non-Markovianity signifies frequent energy or information exchanges, rendering the model closer to reality. There are two limiting cases of interest, the Markovian and non-Markovian regimes.  In the non-Markovian limit ($\gamma\rightarrow0$), the steady-state AEN is found to be proportional to the parameter $\gamma$.  However, in the Markovian limit ($\gamma\rightarrow\infty$), the system reaches a steady state with a temperature identical to that of the bath, as expected. Furthermore, the effectiveness of pulse control for the protection of quantum states was explored in these regimes. The importance of non-Markovianity is shown explicitly in this model. In scenarios involving detuning between the system frequency and the central frequency of the environment, a non-Markovian interaction reveals that the AEN persists in its initial state for a prolonged duration.  This shows that employing pulse control can significantly shield the quantum system from decoherence effects in this situation. Furthermore, the pulse not only extends the relaxation time of the oscillator but can also expedite the relaxation process depending on the circumstances. Lastly, a Mpemba effect can be induced in the non-Markovian regime by implementing a kick pulse.  This is another interesting phenomena that can be observed in this exactly solvable model of an open quantum system.  Such models significantly enhance our understanding of open quantum system evolution by providing examples of interesting phenomena and how they occur in different types of environments.

\section{Acknowledgement}

This work was supported by the National Natural Science Foundation of China (NSFC) under Grants No. 12075050, 12205037, the Natural Science Foundation of Shandong Province (Grant No. ZR2024MA046, ZR2021LLZ004),  the grant PID2021-126273NB-I00 funded by MCIN/AEI/10.13039/501100011033, and by “ERDF A way of making Europe” and the Basque Government through grant number IT1470-22.  This project has also received support from the Spanish Ministry for Digital Transformation and of Civil Service of the Spanish Government through the QUANTUM ENIA project call - Quantum Spain, EU through the Recovery, Transformation and Resilience Plan – NextGenerationEU within the framework of the Digital Spain 2026.MSB was supported by the US National Science Foundation's IR/D program. L.-A. Wu thanks Dr. Qiong-Yi He for early communication on this paper and Dr. D. Segal for helpful discussions.

\clearpage

\appendix
\begin{widetext}
\section{Supplemental Material}		

For the two-order complex coefficient linear homogeneous differential equation
\begin{equation}
	\ddot{A(t)}+(a+ib) \dot{A(t)}+(c+id) A(t)=0.
	\label{eq14}	
\end{equation}
the solution is

1)  Let $\Delta_1=ab-2d, \lam_1=\frac{4c+b^2-a^2}{2(ab-2d)}$, if $c\neq (a^2-b^2)/4$, $d\neq ab/2$,

When $\Delta_1>0$,

\begin{eqnarray}
	A(t)&=&C_1e^{[-\frac{a}{2}+\frac{\sqrt{\Delta_1}}{2}\sqrt{\sqrt{\lambda_1^2+1}-\lambda_1}-\frac{1}{2}(b-\frac{\sqrt{\Delta_1}}{\sqrt{\sqrt{\lambda_1^2+1}-\lambda_1}})i]t}+C_2e^{[-\frac{a}{2}-\frac{\sqrt{\Delta_1}}{2}\sqrt{\sqrt{\lambda_1^2+1}-\lambda_1}-\frac{1}{2}(b+\frac{\sqrt{\Delta_1}}{\sqrt{\sqrt{\lambda_1^2+1}-\lambda_1}})i]t}.
	\label{eq15}
\end{eqnarray}

When $\Delta_1<0$,

\begin{eqnarray}
	A(t)&=&C_1e^{[-\frac{a}{2}-\frac{\sqrt{-\Delta_1}}{2}\sqrt{\sqrt{\lambda_1^2+1}-\lambda_1}-\frac{1}{2}(b-\frac{\sqrt{-\Delta_1}}{\sqrt{\sqrt{\lambda_1^2+1}-\lambda_1}})i]t}+C_2e^{[-\frac{a}{2}+\frac{\sqrt{-\Delta_1}}{2}\sqrt{\sqrt{\lambda_1^2+1}-\lambda_1}-\frac{1}{2}(b+\frac{\sqrt{-\Delta_1}}{\sqrt{\sqrt{\lambda_1^2+1}-\lambda_1}})i]t}.
	\label{eq16}
\end{eqnarray}

2) If $b\neq0$, $d=\frac{ab}{2}$, let $\Delta_2=4c+b^2-a^2$

When $\Delta_2=0$,
\begin{eqnarray}
	A(t)&=&(C_1t+C_2)e^{-(\frac{a}{2}+\frac{b}{2}i)t}.
	\label{eq17}
\end{eqnarray}

When $\Delta_2>0$,
\begin{eqnarray}
	A(t)&=&C_1e^{-(\frac{a}{2}+\frac{-b+\sqrt{\Delta_2}}{2}i)t}+C_2e^{-(\frac{a}{2}+\frac{b+\sqrt{\Delta_2}}{2}i)t}.
	\label{eq18}
\end{eqnarray}

When $\Delta_2<0$,

\begin{eqnarray}
	A(t)&=&C_1e^{(\frac{-a+\sqrt{-\Delta_2}}{2}-\frac{b}{2}i)t}+C_2e^{-(\frac{a+\sqrt{-\Delta_2}}{2}+\frac{b}{2}i)t}.
	\label{eq19}
\end{eqnarray}

3) If $d\neq0$, $c=\frac{a^2-b^2}{4}$, let $\Delta_3=ab-2d$,

When $\Delta_3=0$,
\begin{eqnarray}
	A(t)&=&(C_1t+C_2)e^{-\frac{a+bi}{2}t}.
	\label{eq20}
\end{eqnarray}

When $\Delta_3>0$,

\begin{eqnarray}
	A(t)&=&C_1e^{(\frac{-a+\sqrt{\Delta_3}}{2}+\frac{-b+\sqrt{\Delta_3}}{2}i)t}+C_2e^{-(\frac{a+\sqrt{\Delta_3}}{2}+\frac{b+\sqrt{\Delta_3}}{2}i)t}.
	\label{eq21}	
\end{eqnarray}

When $\Delta_3<0$,

\begin{eqnarray}
	A(t)&=&C_1e^{-(\frac{a-\sqrt{-\Delta_3}}{2}+\frac{b+\sqrt{-\Delta_3}}{2}i)t}+C_2e^{(\frac{-a-\sqrt{-\Delta_3}}{2}+\frac{-b+\sqrt{-\Delta_3}}{2}i)t}.
	\label{eq22}
\end{eqnarray}

For the two-order complex coefficient linear nonhomogeneous differential equation,
\begin{eqnarray}
	\ddot{B_j(t)}+(a+ib) \dot{B_j(t)}+(c+id) B_j(t)=-g_j\left(\omega_{j}-\omega_{0}+i\,\gamma\right)e^{-i\omega_j t}.
	\label{eq23}	
\end{eqnarray}
The solution is the same with Eq.~(\ref{eq14}) for the homogeneous part, except for the initial condition $B_j(t=0)=0,\dot{B_j}(t=0)=-ig_j$. To solve the nonhomogeneous part, we let let $f=-g_j\left(\omega_{j}-\Omega+i\,\gamma\right),\lambda=-i\omega_j$, $p_1=a+bi, p_2=c+di$, the characteristic equation is
\begin{eqnarray}
	\Phi(r)=r^2+p_1 r+p_2.
\end{eqnarray}
The special solution is

1) if $\Phi(\lambda)\neq0$,
\begin{eqnarray}
	B_j^{\prime}(t)=\frac{f}{\Phi(\lambda)}e^{\lambda t}.
\end{eqnarray}
2) if $\Phi(\lambda)=0$, $\dot{\Phi(\lambda)}\neq0$

\begin{eqnarray}
	B_j^{\prime}(t)=\frac{f}{\dot{\Phi(\lambda)}}te^{\lambda t}.
\end{eqnarray}

\end{widetext}
\end{document}